\documentclass[twocolumn,amsmath,amsfonts,amssymb,aps,prx,preprintnumbers,superscriptaddress,longbibliography]{revtex4-2}

\usepackage[utf8]{inputenc}
\usepackage[T1]{fontenc}
\usepackage{lmodern}
\usepackage{graphicx}
\usepackage{dcolumn}
\usepackage{bm}
\usepackage{textcomp}
\usepackage{float}

\usepackage[normalem]{ulem}
\usepackage{ifpdf}
\usepackage[squaren,Gray]{SIunits}

\usepackage{amsmath}

\ifpdf
\usepackage{epstopdf}
\usepackage[pdftex,unicode,pdfstartview={FitH},pdfborder={0 0 0}]{hyperref}
\usepackage{hypcap}
\else
\usepackage[hypertex]{hyperref}
\fi
\hypersetup{
    bookmarksnumbered = true,
    colorlinks = true, linkcolor = darkblue,
    citecolor = darkblue, filecolor = darkblue,
    menucolor = darkblue, urlcolor = black
}

\usepackage{color}
\definecolor{red}{rgb}{1,0,0}
\definecolor{blue}{rgb}{0.2,0,0.8}
\definecolor{black}{rgb}{0,0,0}

\usepackage{color}
\definecolor{darkblue}{rgb}{0,0,0.7}

\definecolor{red}{rgb}{1,0,0}

\definecolor{green}{rgb}{0,0.6,0}

\definecolor{grey}{rgb}{0.7,0.7,0.7}

\definecolor{orange}{rgb}{0.8,0.4,0}

\usepackage{titlesec}           
\titleformat{\subsection}
{\bfseries} 
{}          
{0.0cm}     
{}          
[]          

\begin{document}

\title{Optical switching of magnetic order in few-layer CrSBr}

\def\LMU{Fakult\"at f\"ur Physik, Munich Quantum Center, and Center for NanoScience (CeNS), Ludwig-Maximilians-Universit\"at M\"unchen, Geschwister-Scholl-Platz~1, D-80539 M\"unchen, Germany}

\author{Lukas Husel}
\altaffiliation{These authors contributed equally to this work}
\affiliation{\LMU}
\affiliation{Department of Physics, University of Basel, 4056 Basel, Switzerland}

\author{Julian Trapp}
\altaffiliation{These authors contributed equally to this work}
\affiliation{\LMU}

\author{Moritz Würf}
\altaffiliation{These authors contributed equally to this work}
\affiliation{\LMU}

\author{Anna Rupp}
\affiliation{\LMU}

\author{Tim Wedl}
\affiliation{\LMU}

\author{Kenji Watanabe}
\affiliation{Research Center for Electronic and Optical Materials, National Institute for Materials Science, 1-1 Namiki, Tsukuba 305-0044, Japan}

\author{Takashi Taniguchi}
\affiliation{Research Center for Materials Nanoarchitectonics, National Institute for Materials Science, 1-1 Namiki, Tsukuba 305-0044, Japan}

\author{Iva Plutnarova}
\affiliation{Department of Inorganic Chemistry, University of Chemistry and Technology Prague, Technická 5, 166 28 Prague 6, Czech Republic}

\author{Zdenek Sofer}
\affiliation{Department of Inorganic Chemistry, University of Chemistry and Technology Prague, Technická 5, 166 28 Prague 6, Czech Republic}

\author{Alexander H\"ogele}
\affiliation{\LMU}
\affiliation{Munich Center for Quantum Science and Technology (MCQST), Schellingstr.~4, D-80799 M\"unchen, Germany}

\begin{abstract}
Manipulating magnetism with light is crucial for both fundamental understanding and technological advancements of information storage devices. The layered antiferromagnet CrSBr, part of the recently emerging class of two-dimensional van der Waals magnets, offers a unique path towards optical control of magnetism via magneto-excitons, which allow optical readout of the spin alignment and also provide a strong absorption channel. Here, we use exciton absorption and photoluminescence to demonstrate optical switching of the magnetic order in bi- and trilayer CrSBr. Using a continuous-wave laser with a power as low as a few microwatts at near-critical external magnetic fields, we demonstrate both local and remote switching of the magnetic configuration in extended lateral domains and further employ this mechanism to deterministically prepare the zero-field magnetic configuration. Our results establish optical switching as a useful means of controlling magnetism in CrSBr, with potential applications in magneto-optoelectronic devices.
\end{abstract}

\maketitle

The optical manipulation of magnetic order is of key significance in solid-state science and device design. Light couples to magnetic moments through both direct, non-absorptive pathways and indirect channels in which absorbed energy alters the magnetic state thermally, magnetoelastically, or via photoexcited carriers. The associated physical mechanisms span a wide range of timescales, from femtosecond spin dynamics to diffusive heating on the order of seconds~\cite{kirilyuk2010}, and enable deterministic optical switching and magnetization control. Ultimately, these effects form the basis for new classes of memory devices with improved write speed and reduced switching energy~\cite{sander2017,kimel2019}, as well as novel functionality such as heat-assisted magnetic recording~\cite{kryder2008}.

Two-dimensional (2D) van der Waals magnets~\cite{gibertini2019,jiang2021,park2026} constitute a unique platform to study the influence of light on the magnetic order, enabling ultrathin devices~\cite{jiang2018,zhang2021,zhang2024} and straightforward integration in micro- and nanophotonic structures for enhanced light-matter interactions~\cite{demir2025,dirnberger2023,wang2023,ziegler2025}. Within this material class, the semiconducting A-type antiferromagnet CrSBr~\cite{wang2020,lee2021,lopez-paz2022,Ziebel2024} constitutes a particularly promising platform: strongly bound excitons coupled to the magnetic order~\cite{wilson2021,bae2022,klein2023,adak2026} provide simple and efficient optical readout of the magnetic configuration \cite{Tabataba-Vakili2024} as well as an efficient absorption channel for near-infrared light. Leveraging these properties for optical magnetization switching, as recently demonstrated in the 2D magnets CrI$_3$~\cite{zhang2022} and Fe$_3$GeTe$_2$~\cite{liu2020}, would provide additional control of magnetism complementary to external magnetic field~\cite{wilson2021}, temperature~\cite{lopez-paz2022}, strain~\cite{cenker2022}, or charge doping~\cite{Tabataba-Vakili2024,jo2024,hong2025}. 

\begin{figure*}[t]
\centering \includegraphics[scale=1.]{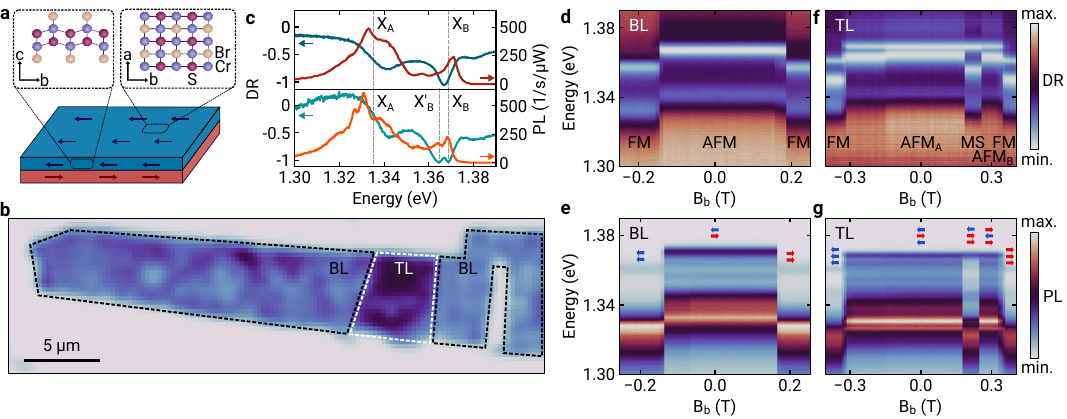}
\caption{\textbf{Excitons in few-layer CrSBr as sensors of magnetic order.} \textbf{a}, Schematic of CrSBr bilayer cooled below the Néel temperature, with the magnetic configuration indicated by the arrows. The insets show the crystal structure in the b-c and a-b planes. \textbf{b}, Confocal PL raster-scan map of CrSBr bilayer (BL) and trilayer (TL) regions encapsulated by hBN. The PL was integrated in the spectral range $1.327-1.332$~eV. \textbf{c}, Top panel: Spectra of bilayer PL (red line) and DR (blue line). Bottom panel: Trilayer PL (orange line) and DR (light green line). X$_\mathrm{A}$, X$_\mathrm{B}$ and X$'_\mathrm{B}$ exciton energies are indicated by the dashed lines. \textbf{d} and \textbf{e}, Bilayer DR and PL spectra as a function of magnetic field B$_\mathrm{b}$ applied along the crystal b-axis and ramped from negative to positive values. The magnetic configuration switches between the ferromagnetic (FM) and antiferromagnetic (AFM) phases as indicated by the arrows. \textbf{f} and \textbf{g}, Same as \textbf{d} and \textbf{e}, but for the trilayer with a metastable (MS) configuration in between the FM and AFM phases.}
\label{fig1}
\end{figure*}

Here, we report optical switching of the magnetic order in few-layer CrSBr. We demonstrate that at near-critical external magnetic fields, low-power continuous-wave (cw) optical excitation induces spin-flip transitions by transferring energy to the spin lattice via exciton absorption. We observe optical switching of extended magnetic domains, as well as remote switching of neighboring domains energetically linked through lateral exchange bias~\cite{pellet-mary2025,sun2025}. Finally, we utilize this mechanism to deterministically define and read out the zero-field magnetic configuration. Our results demonstrate that the switching behavior, as determined from photoluminescence (PL) experiments, critically depends on the pump laser power and establish optical switching as a resource for magnetization control in CrSBr magneto-optical devices. 

In our work, we investigated bi- and trilayer CrSBr encapsulated by hexagonal boron nitride (hBN) and cooled to $4$~K, below the Néel temperature. At low temperatures and in the absence of an external magnetic field, the ground-state magnetic configuration is ferromagnetic (FM) within each layer, while adjacent layers exhibit antiferromagnetic (AFM) order~\cite{lee2021}, as illustrated in Fig.~1a for a CrSBr bilayer. The hard and easy magnetic axes are aligned with the c- and b-axis of the crystal, respectively. A raster-scan map of confocal PL intensity is shown in Fig.~1b for the relevant area of a CrSBr flake with a trilayer region bounded by two bilayer regions.

CrSBr hosts strongly bound excitons with near-infrared optical transitions~\cite{wilson2021,klein2023,Tabataba-Vakili2024,smiertka2026} conveniently probed by confocal spectroscopy. On the bilayer region, the differential reflectivity (DR) spectrum (blue line in the top panel of Fig.~1c) features signatures of X$_\mathrm{A}$ and X$_\mathrm{B}$ excitons with respective energies of $1.335$ and $1.369$~eV, consistent with our previous observation on hBN-encapsulated few-layer samples~\cite{Tabataba-Vakili2024}. The corresponding PL spectrum is shown by the red line in the top panel of Fig.~1c. Trilayer DR and PL (green and orange lines in the bottom panel of Fig.~1c) exhibit an additional resonance X$'_\mathrm{B}$, stemming from excitons in the middle CrSBr layer and redshifted from X$_\mathrm{B}$ by $5$~meV~\cite{Tabataba-Vakili2024}. 

The response to an external magnetic field applied along the b-axis is inferred from DR and PL spectra, as shown for the bilayer in Figs.~1d and e. At critical field values B$_\mathrm{crit}$, the magnetic configuration switches from AFM to FM via a spin-flip transition, which enables interlayer carrier tunneling and consequently lowers the exciton energy through interlayer hybridization~\cite{wilson2021}. While the bilayer comprises only AFM and FM phases, the trilayer features a richer response, with DR and PL shown in Figs.~1f and g. Exciton spectroscopy, combined with a theoretical analysis~\cite{Tabataba-Vakili2024,pellet-mary2025}, has established the sequence of magnetic configurations indicated by the arrows in Fig.~1g: at positive fields, the system switches between the two possible AFM configurations, referred to as AFM$_\mathrm{A/B}$, via a metastable state MS. 

\begin{figure}[t]
    \centering \includegraphics[scale=1]{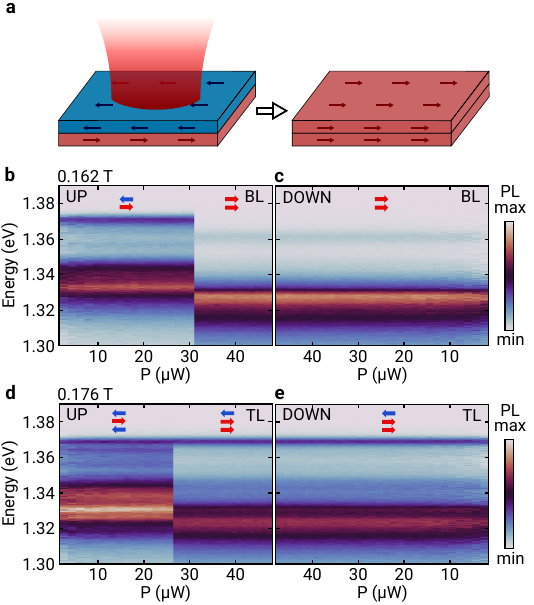}
    \caption{\textbf{Optical switching of magnetic spin-flip transitions.} \textbf{a}, Schematic of the experiment: A non-resonant cw laser switches the magnetic configuration in few-layer CrSBr with illumination powers as low as a few tens of microwatts. \textbf{b}, Bilayer PL as a function of laser power (upward sweep) in a magnetic field of $0.162$~T applied along the b-axis. The magnetic configuration switches from AFM to FM, as indicated by the arrows. \textbf{c}, Same for a decreasing laser power (downward sweep), performed directly after the upward sweep in \textbf{b}. \textbf{d} and \textbf{e}, Same as \textbf{b} and \textbf{c}, but for the trilayer in a magnetic field of $0.176$~T with laser-induced AFM$_\mathrm{A}$--MS spin-flip transition. All spectra were normalized to the laser power.}
    \label{fig2}
\end{figure}

\begin{figure*}[t]
    \centering \includegraphics[scale=1.]{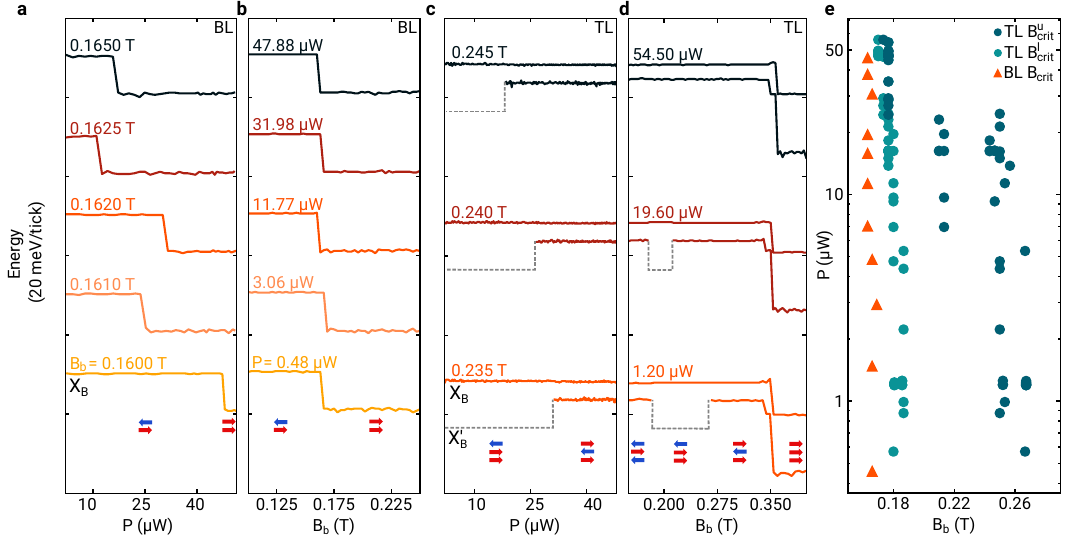}
    \caption{\textbf{Interplay of optical and magnetic fields determines the switching behavior.} \textbf{a}, PL energy of the X$_\mathrm{B}$ exciton in the bilayer, measured for upward sweeps of the laser power at different magnetic fields B$_\mathrm{b}$. The arrows depict the magnetic configuration. \textbf{b}, Bilayer PL energy of X$_\mathrm{B}$ for upward sweeps of B$_\mathrm{b}$, measured at different laser powers. \textbf{c} and \textbf{d}, Same as \textbf{a} and \textbf{b}, but for the trilayer. In the MS magnetic configuration, the X$’_\mathrm{B}$ exciton PL was quenched (the dashed lines indicate its energy in DR). \textbf{e}, Critical spin-flip fields as a function of the laser power. For the bilayer, B$_\mathrm{crit}$ for the AFM--FM transition is shown by orange triangles; for the trilayer, the critical fields for the AFM$_\mathrm{A}$--MS and MS--AFM$_\mathrm{B}$ (B$_\mathrm{crit}^{\mathrm{l}}$ and B$_\mathrm{crit}^{\mathrm{u}}$, respectively) are shown by green and blue dots.}
    \label{fig3}
\end{figure*}

Surprisingly, in addition to these characteristic signatures of magnetic transitions controlled by the magnetic field, we observed optically induced spin-flip transitions indicated schematically in Fig.~2a. Our scheme initializes the left bilayer in Fig.~1b in the AFM configuration and subjects it to a magnetic field just below the critical value B$_\mathrm{crit} = 0.1680$~T. The left panel in Fig.~2b shows the PL evolution as a function of power of a non-resonant cw excitation laser at $1.450$~eV. Just above $30$~µW, the magnetic configuration switches from AFM to FM, as signified by the redshift in the exciton energy. Upon decreasing the excitation power back to zero, the system remains in the FM phase, as shown in Fig.~2c. We repeated this measurement sequence on the trilayer for the AFM$_\mathrm{A}$--MS transition (B$_\mathrm{crit} = 0.185$~T) and found similar results, with data in Figs.~2d and e. Evidently, sufficiently high laser power facilitates switching of the magnetic configuration.

To study this intricate interplay of optical switching and magnetic field, we performed a systematic analysis of bi- and trilayer regions, with data shown in Fig.~3. For the bilayer, Fig.~3a shows the X$_\mathrm{B}$ exciton PL energy as a function of increasing laser power, for different values of B$_\mathrm{b}$ close to the spin-flip transition field B$_\mathrm{crit} = 0.1680$~T. Minimizing the difference $\Delta = \mathrm{B}_\mathrm{crit} - \mathrm{B}_\mathrm{b}$ between applied and critical fields minimizes the required switching power P$_\mathrm{crit}$ (top and bottom traces in Fig.~3a). Similar evidence holds for the trilayer MS--AFM$_\mathrm{B}$ transition, with B$_\mathrm{crit} = 0.250$~T and data shown in Fig.~3c. Comparing the bi- and trilayer behavior at similar values of $\Delta$, we find a much lower P$_\mathrm{crit}$ for the trilayer, suggesting a lower energy cost for switching of the metastable configuration.

We now exchange the roles of magnetic field and laser power, sweeping the former while keeping the latter constant, with data for the bi- and trilayer shown in Figs.~3b and d, respectively. While the bilayer critical field changed only marginally at laser powers up to $50$~µW, the trilayer MS--AFM$_\mathrm{B}$ critical field decreased considerably with increasing power. A systematic measurement of the critical field as a function of the laser power, with data shown in Fig.~3e, reveals three distinct switching regimes for the trilayer MS--AFM$_\mathrm{B}$ transition. At low powers, the critical magnetic field shows no significant power dependence. In an intermediate power regime of $7-12$~µW, two distinct critical-field values emerge, with switching occurring randomly at either value in a single field-sweep sequence. This behavior suggests an additional switching path which can become energetically favorable in the presence of optical excitation. At powers above $12$~µW, the MS--AFM$_\mathrm{B}$ critical field collapses to that of the AFM$_\mathrm{A}$--MS transition. High optical powers render the MS unstable and eliminate its spectral signatures as in the top trace in Fig.~3d.

By expanding our analysis beyond individual crystal positions, we found that optical switching affects the magnetic configuration across laterally extended areas. We first focus on the left bilayer in Fig.~1b. By increasing the magnetic field to values slightly above the critical field, we observed switching of a sub-region of the bilayer, as evidenced by the different PL intensities in the maps of Figs.~4a and b. The laser power in these measurements was kept sub-critical to avoid optical switching. The fact that only a sub-domain of the bilayer switched at the given magnetic field is likely related to magnetization-pinning by local disorder. After a re-initialization of the bilayer in the AFM configuration, we applied a near-critical field and illuminated the spot shown by the red dot in Fig.~4c with a laser power above P$_\mathrm{crit}$. The PL map in Fig.~4d, obtained subsequently at sub-critical power, shows that local laser illumination switched the same extended domain of the crystal as in response to a global over-critical magnetic field. 

\begin{figure*}[t]
    \centering \includegraphics[scale=1.]{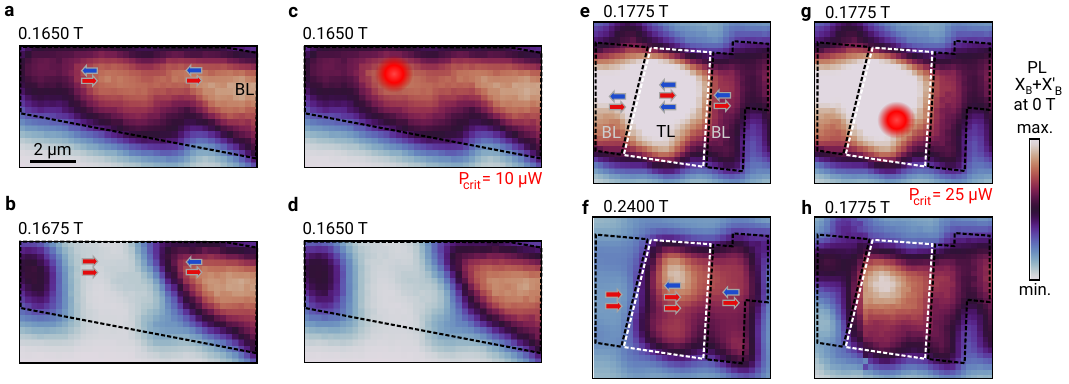} 
    \caption{\textbf{Local and remote switching of magnetic domains.} \textbf{a} and \textbf{b}, Raster-scan maps of bilayer PL intensity for magnetic fields of $0.1650$ and $0.1675$~T, respectively, acquired for a laser power below the critical value P$_\mathrm{crit}$. The increase in magnetic field switches the magnetic configuration within a subdomain, as indicated by the arrows. \textbf{c} and \textbf{d}, PL map of the same area as in \textbf{a} and \textbf{b} at a constant magnetic field of $0.1650$~T, acquired before and after illuminating the spot indicated by the red dot with a laser power above P$_\mathrm{crit}$. Local laser illumination has the same effect as an increase of the magnetic field. \textbf{e} and \textbf{f}, PL map of the trilayer and the neighboring bilayers for magnetic fields of $0.1775$ and $0.2400$~T, respectively, acquired for a laser power below P$_\mathrm{crit}$. Here, increased magnetic field switched both the left bilayer and the trilayer. \textbf{g} and \textbf{h}, PL map of the same region as in \textbf{e} and \textbf{f}, acquired before and after illuminating the trilayer with a laser power above P$_\mathrm{crit}$. In all panels, the PL intensity was integrated in the spectral range of zero-field X$_\mathrm{B}$ and X$'_\mathrm{B}$ energies.}
    \label{fig4}
\end{figure*}

\begin{figure}[t]
    \centering \includegraphics[scale=1]{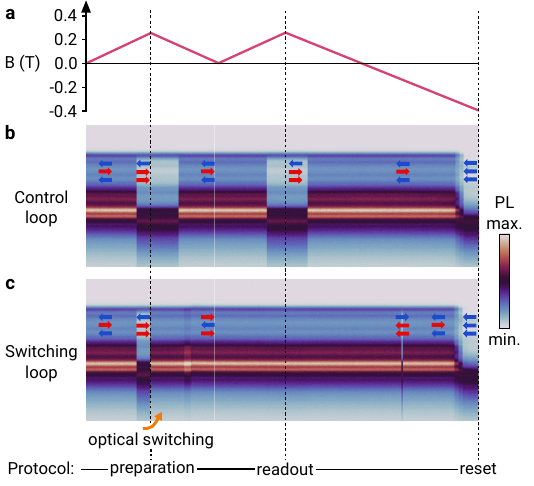} 
    \caption{\textbf{Magneto-optical preparation and detection of the zero-field magnetic configuration.} \textbf{a}, Sequence of magnetic fields B$_\mathrm{b}$ applied in the experiment. \textbf{b}, PL spectra obtained for the magnetic field sequence in \textbf{a} with a laser power below P$_\mathrm{crit}$, corresponding to a control measurement loop. The magnetic configurations occurring during the sweep are indicated by the arrows. \textbf{c}, Same as \textbf{b}, but with the sequence modified by targeted optical switching as indicated by the arrow.}
    \label{fig5}
\end{figure}

We repeated the measurement sequence on the trilayer region, focusing on the AFM$_\mathrm{A}$--MS transition. For small laser powers, increasing the magnetic field switched both the left bilayer and the trilayer. This connected switching is likely a result of lateral exchange bias between the different multilayer regions~\cite{pellet-mary2025, sun2025}, which acts as an effective in-plane magnetic field bias for the energetic cost of domain flips. This has intriguing consequences for the optical switching, which we demonstrate by illuminating the trilayer at the position indicated in Fig.~4g with laser powers above P$_\mathrm{crit} = 25$~µW. Laser illumination not only switched the full trilayer, as already observed in the isolated bilayer region, but also remotely switched the connected bilayer. Vice versa, optical switching of the bilayer also induced the switch of the full trilayer region (data not shown). We also observed a similar effect for the AFM$_\mathrm{B}$--FM transition: here, the trilayer and the right bilayer regions exhibited spin-flip transitions at the same critical field, and both regions switched as one of them was illuminated with laser powers above P$_\mathrm{crit}$. 

Evidently, the rich phenomena of optically induced magnetic spin-flip transitions are intimately related to the photogeneration of excitons. This assignment is supported by our observation that one order of magnitude lower switching powers were required when the laser was tuned into resonance with the X$_\mathrm{B}$ exciton transition. Consistent with this finding, we exclude the trivial effect of substrate heating due to the absence of optical switching under laser illumination cross-polarized to the dipole axis of the quasi one-dimensional excitons~\cite{klein2023} (data not shown). Moreover, the absence of trion signatures~\cite{Tabataba-Vakili2024} and thermal shifts~\cite{dirnberger2026}, as evidenced by constant exciton energy and linewidth for all laser powers in Figs.~2b and d, excludes sizable photo-doping or crystal heating. We therefore argue that phonons~\cite{lin2024} and magnons~\cite{diederich2023,dirnberger2026}, launched by the excitation and relaxation of excitons, provide the energy required for switching the local spin-lattice configuration without significant heating effects.

Finally, we demonstrate how optical switching of trilayer CrSBr can be employed to deterministically prepare and read out the zero-field magnetic configuration, similar to a magnetic memory bit. We initialize the AFM$_\mathrm{A}$ state and apply the sequence of magnetic fields shown in Fig.~5a. A control measurement, performed at sufficiently low laser powers to prevent optical switching,  demonstrates in Fig.~5b reversible access to the MS spin state at $0.23$~T, which, at positive fields, heralds the zero-field magnetic configuration of AFM$_\mathrm{A}$. In Fig.~5c, we now modify this sequence by using an overcritical laser power to optically switch the MS state into the AFM$_\mathrm{B}$ state. This process is irreversible upon reduction of the magnetic field, such that the zero-field configuration is deterministically set to AFM$_\mathrm{B}$. This can be confirmed non-invasively via the detection of the MS state, which appears at positive fields only if the zero-field configuration is AFM$_\mathrm{A}$, and therefore is absent in Fig.~5c as the field is ramped up to $0.23$~T. Such sequences should thus enable binary information storage in magnetic configurations with read-out via the MS state. 

In conclusion, we have demonstrated optical switching of the magnetic order in bi- and trilayer CrSBr. At near-critical magnetic fields, we deterministically induced spin-flip transitions in extended lateral areas using a cw laser with a power of a few microwatts. This mechanism also enabled remote switching of neighboring magnetic domains in a connected bi-trilayer system. Moreover, by selectively manipulating the magnetic hysteresis sequence, we demonstrated deterministic preparation and non-invasive readout of the zero-field magnetic configuration. Related schemes could facilitate independent control of non-connected crystal areas~\cite{sun2025} within a single device, or writing of domain walls in lateral-exchange-biased CrSBr multilayer systems~\cite{pellet-mary2025}. Beyond these implications, we anticipate that time-resolved PL experiments, performed across larger device areas, will offer valuable insight into the switching dynamics. We anticipate that optical switching can be extended to field-free, all-optical operation by exploiting the recently reported twist-tuned hysteresis in CrSBr~\cite{Mondal2026}. These perspectives establish optical switching as a resource for fundamental studies and applications of two-dimensional magnetic semiconductors.\\

\noindent \textbf{METHODS}

\noindent \textbf{Sample fabrication:} Standard exfoliation was employed to obtain few-layer CrSBr from bulk crystals synthesized by chemical vapor transport as well as hBN flakes. The van der Waals heterostack consisting of few-layer CrSBr sandwiched between two hBN layers (with bottom and top thickness of $54$ and $16$~nm, respectively) was assembled using a dry-transfer process based on hemispherical polydimethylsiloxane droplets and poly(bisphenol A-carbonate) polymer films with thicknesses ranging from $1.5$~µm to $3$~µm. Typical pick-up temperatures were in the range $110-130$~°C, the release temperature was $200$~°C. The heterostack was placed on a distributed Bragg reflector with the stopband center wavelength of $900$~nm (Qlibri GmbH). The bottom hBN thickness was chosen to place the CrSBr crystal close to an antinode of the optical field reflected from the mirror~\cite{rogers2020, horng2020}.

\noindent \textbf{Cryogenic spectroscopy:} The experiments were performed in backscattering geometry inside a closed-cycle magneto-cryostat (attocube systems, attoDRY1000), with the sample mounted in Voigt geometry to apply the magnetic field along the b-axis of the CrSBr crystal. An aspheric lens (Thorlabs 354330-B, NA=0.68) was used to focus the light onto the sample and to collect confocally the white-light reflection or PL into a single-mode fiber connected to a spectrometer (Roper Scientific Acton SP2500 with Spec-10:100BR). For DR measurements, obtained by normalizing the reflection spectrum ($R$) from the heterostack region by the spectrum from the sample region without the heterostack ($R_0$) as $\textrm{DR} = (R-R_0)/R_0$, a halogen lamp (Ocean Optics) was employed. Non-resonant and resonant cw PL excitation were performed using a laser diode (Roithner QL856JS-AL) and a power-stabilized titanium sapphire laser (MSquared SolsTiS), respectively. The linear polarization of the excitation beam was aligned with the crystal b-axis unless otherwise stated. The PL map in Fig.~1b was obtained during characterization in another cryostat (attocube systems, attoDRY800) using a similar optical setup with the aspheric lens replaced by a cryogenic apochromatic objective (attocube systems, LT-APO). To avoid heating effects caused by magnetic field ramps with ramp speeds of $1$~mT/s and step sizes between $2.5$ and $10$~mT, waiting times were introduced prior to recording the spectra with integration times between $0.1$ and $1$~s.\\

\noindent \textbf{ACKNOWLEDGEMENTS}\\  
\noindent We gratefully acknowledge helpful discussions with C.~ Pellet-Mary, P.~Knüppel, K.~Wagner, D.~Dutta, P.~Maletinsky, and F.~Dirnberger. This research was funded by the Deutsche Forschungsgemeinschaft (DFG, German Research Foundation) under Germany's Excellence Strategy EXC-2111-390814868 (MCQST). A.\,R. and L.\,H. acknowledge funding by the Hightech Agenda Bayern Plus within the Munich Quantum Valley (MQV) and EQAP funding initiatives. K.\,W. and T.\,T. acknowledge support from the JSPS KAKENHI (Grant Numbers 21H05233 and 23H02052), the CREST (JPMJCR24A5), JST and World Premier International Research Center Initiative (WPI), MEXT, Japan. Z.\,S. was supported by the ERC-CZ program (project LL2101) from the Ministry of Education, Youth, and Sports (MEYS), and by the Advanced Multiscale Materials for Key Enabling Technologies project, supported by the Ministry of Education, Youth, and Sports of the Czech Republic, Project No. CZ.02.01.01/00/22\_008/0004558 and co-funded by the European Union.\\

\noindent \textbf{AUTHOR CONTRIBUTIONS}\\ 
\noindent M.\,W. and T.\,W. fabricated heterostacks from CrSBr crystals provided by I.\,P. and Z.\,S. as well as hBN crystals provided by T.\,T. and K.\,W.. L.\,H., J.\,T. and M.\,W. performed the experiments presented in this work, with complementary experiments performed by A.\,R.. L.\,H., J.\,T., M.\,W. and A.\,H. analyzed the data and wrote the manuscript. L.\,H., J.\,T. and M.\,W. contributed equally to this work.\\

\noindent \textbf{CORRESPONDING AUTHORS} \\
\noindent lukas.husel@physik.lmu.de, alexander.hoegele@lmu.de\\

\noindent \textbf{DATA AVAILABILITY}
  
\noindent The data that support the findings of this study are available from the corresponding authors upon request. \\

\noindent \textbf{COMPETING INTERESTS}  

\noindent The authors declare no competing interests.
\bibliography{PaperLiteratur}

\end{document}